\documentclass{mem}
\usepackage{graphicx}
\usepackage{times}
\usepackage{balance}
\usepackage{natbib}
\usepackage{txfonts}
\usepackage[a4paper]{hyperref}
\begin{document}

\title{Cosmological Shocks in Eulerian Simulations: Main Properties 
and Cosmic Rays Acceleration}

 \subtitle{}
\author{
F.Vazza\inst{1 \and 2}, G. Brunetti\inst{2} \and  C. Gheller\inst{3}}
\offprints{F.Vazza, \email{vazza@ira.inaf.it}}
\institute{
Dipartimento di Astronomia, Universit\'a di Bologna, via Ranzani
1,I-40127 Bologna, Italy  \and INAF/Istituto di Radioastronomia, via Gobetti 101, I-40129 Bologna \and  CINECA, High Performance System Division, Casalecchio di
Reno--Bologna, Italy}

%\date{Accepted ???. Received ???; in original form ???}

\abstract{{\it Aims}: morpholgies, number and energy distributions
 of Cosmological Shock Waves from a set of ENZO cosmological simulations 
are produced, along with a study of the connection with Cosmic Rays processes
 in different environments.
{\it Method:} we perform cosmological simulations with the public release of 
the PPM code ENZO, adopt a simple and physically motivated numerical setup 
to follow the evolution of cosmic structures at the resolution of $125kpc$ 
per cell, and characterise shocks with a new post processing scheme.
{\it Results:} we estimate the efficency of the acceleration of Cosmic Ray
particles and present the first comparison of our results with existing
limits from observations of galaxy clusters. 

%% ======================================================================= %%

}

\titlerunning{Shocks in ENZO}

\authorrunning{Vazza, Brunetti \& Gheller}

\maketitle

\section{Introduction}

\label{sec:intro}

Galaxy clusters store up to several $10^{63}$ ergs in the form of hot
baryonic matter, due to the action of shock-heating processes in large
scale structures formation \citep{ze70}.
Detecting shocks in Large Scale Structures (LSS) is still 
observationally challenging since they should more frequently develop 
in peripheral, low X emitting regions of clusters, due to the drop in 
the sound speed there (e.g. \citealt{mv07}).
Shocks are important not only to understand the heating of the ICM  
but also because they are expected to be efficient accelerators of  
supra--thermal particles (e.g. \citealt{sa99}), which can then be advected 
and accumulated inside galaxy clusters (e.g. \citealt{vab96}, \citealt{bbp97}). 
Non thermal activity in galaxy clusters related to 
the presence of Cosmic Ray (CR) electrons and $\sim \mu G$ magnetic field is
proved by radio observations,
which show synchrotron emission in a fraction of merging clusters (e.g. \citealt{fe05}),  in form
of Radio Haloes (at the cluster center) and 
Radio Relics (elongated and at the cluster periphery). 
CR protons are expected to give $\gamma$--Ray emissions from galaxy clusters via the
decay of $\pi^{0}$ generated during proton--proton collisions in the intra galactic
medium. Still, only upper limits of this $\gamma$--Ray emissions have been obtained
so far (e.g. \citealt{re03}).
Recent numerical works claimed that an efficient CR protons acceleration
can occur in large scale shocks \citep{mi01,ry03,pf06}.
However, the identification 
and characterisation of shocks, 
as well as the calculation of the energy injected 
in the form of CR, remain challenging due to the
uncertainties in the numerical schemes and due to our ignorance
of the efficency of the acceleration of CRs at relatively weak shocks. 

In this paper we follow the approach of the seminal paper
by \citet{ry03}, studying the shock wave patterns 
in LSS. Shocks are characterised in a post--processing
phase with a new scheme, which evaluates shocks Mach 
number by analysing the jumps in the velocity varaible.
Estimates of the level of CR injections at these shocks
are provided and compared to present day observational upper limits.
More detailed discussions and presentations of the results
can be found in Vazza, Brunetti \& Gheller (2008 submitted to MNRAS, 
hereafter VBG).

\section{Numerical Code - ENZO.}
\label{sec:enzo}

The simulations described in this paper were performed with
the cosmological code ENZO (e.g. \citealt{osh04}). 
ENZO couples
an N-body particle-mesh solver for Dark Matter  with an adaptive mesh method for ideal  fluidynamics. The hydrodynamical solver is
based on the the Piecewise Parabolic 
Method (PPM, \citealt{cw84}), 
which is a higher order extension of Godunov's shock capturing
method. It is at least second--order accurate in space (up
to the fourth--order, in the case of smooth flows and small timesteps) and
second--order accurate in time, and it is thus a highly suitable hydro method
to study shock pattern. 

\section{Cosmological Simulations.}

We adopt the  "Concordance" model, with
density parameters $\Omega_0 = 1.0$, $\Omega_{BM} = 0.044$, $\Omega_{DM} =
0.226$, $\Omega_{\Lambda} = 0.73$, Hubble parameter $h = 0.71$, 
a power spectrum
produced according to the \citet{eh99} fitting formulas with 
a primordial spectrum normalization $\sigma_{8} = 0.94$, and an initial
redshift of $z=50$.
In order to have a large number stastics of massive
galaxy clusters, we collect several boxes in order to produce a final
 equivalent
volume of $\approx (100Mpc/h)^{3}$ at the fixed numerical 
resolution of 125 kpc,
Our fiducial model here is 
an ensamble of non-radiative simulations with a
a post-processing treatment of 
cosmic reionizaation, designed to reproduce the \citet{hm96}  model.

\section{The Velocity--Jump Method}

\label{sec:algo}

The crossing of a shock in a simulated volume leaves its imprint as 
a jump in all the thermodynamical variables, which can be inverted to
evaluate the shock Mach number, $M$, by means of the standard Rankine--Hugoniot
jump conditions.
The relationship between the jumps in the velocity field, $\Delta v$, 
and $M$ for an idealized shock wave running in an imperturbed medium 
follows from the conservation of momentum and density across 
the shock, and transforming the velocities from the shock frame 
to the Lab frame we get:

\begin{equation}
\Delta v =\frac{3}{4}v_{s}\frac{1-M^{2}}{M^{2}}.
\label{eq:mach_v}
\end{equation}

where $v_{s} = M c_{s}$ and $c_{s}$ is the sound velocity computed
in pre--shocked cells.

The procedure adopted is the following: 
1) we consider only ''candidate'' shocked cells 
with a negative 3--D velocity divergence; 2) in the case of two adiacent
candidate shocked cells, the one with the minimum 3--D divergence is considered
as the post--shock region; 3) we perfrom 1--D scan 
along each axes measuring all $\Delta v_{x,y,z}$
jumps across neighbours cells; 4) we measure the shock Mach number 
along each coordinate according to Eq.\ref{eq:mach_v}, where $v_{s}$ is
calculated from the temperature of the pre--shock cell; 5) the
total Mach number for the shocked cell is finally calculated as 
$M = (M_{x}^{2}+M_{y}^{2}+M_{z}^{2})^{1/2}$.

As in other methods relying on a post-processing of the simulated output, 
this method has its major source of uncertainty in the assumption
of an unperturbed 
velocity field prior to the passage of the shock. 
However, in the case of cosmological simulations
the situation is more complex due to the chaotic pattern of
velocity and temperature fluctuations which 
develops during LSS formation 
(e.g. \citealt{do05}). 
The uncertainties deriving from that are discussed in VBG, where we
claim that in the case of dense cosmological regions the efficency
of the Velocity Jump (VJ) method in characterising shocks is similar to
that of methods based on temperature jumps across cells within clusters,
and becomes better in lower density regions (i.e. filaments and voids).

\begin{figure}
\includegraphics[width=0.49\textwidth]{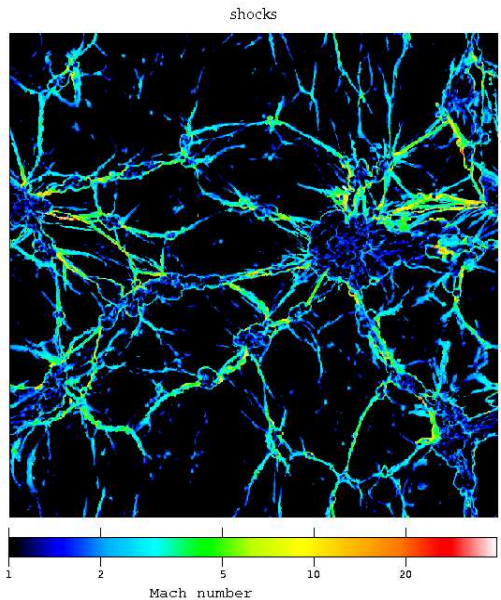}
\caption{2--dimensional slice of the reconstructed Mach numbers of 
shocks for a box of side $80Mpc$.}
\label{fig:mappe}
\end{figure}

\section{RESULTS}
\label{sec:results}

In this Section we present the main results obtained for
shocks in LSS at $z=0$ with the VJ method. 
Additional 
results concerning the time evolution of
shocks in the cosmological volume can be found in VBG.

\begin{figure}
\includegraphics[width=0.49\textwidth]{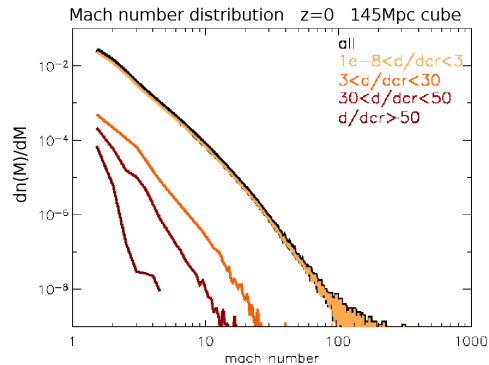}
\caption{Distribution of shocks Mach number for the whole fiducial data--set (black line)
and distributions for subsamples in the cosmic environment (colour lines).}
\label{fig:histo}
\end{figure}

\subsection{Maps and Number distributions.}
\label{subsec:machs}

Fig.\ref{fig:mappe} shows the distribution of detected shocks in a $125kpc$
cut of a region of side $80Mpc$, taken at $z=0$.
Roughly a $\sim 15$ per cent of the simulated volume
hosts shocks at present epoch, with
the percentage of shocked cells increasing in denser environments.
Overall, the picture is pretty similar to early 
results by \citet{mi00},
whith filamentary and sheet--like shocks developing at the interfaces of 
cosmic filaments and sheets, and with shocks surrounding galaxy clusters 
showing spherically shaped boundaries 
at a typical distance of about $1-2 R_{vir}$ from clusters center.
Internal merger shocks are more
irregular and weak, $M \leq 2$, while slightly 
stronger shocks 
are only episodically found within clusters in case of merger events.
An issue which is still poorly addressed in the literature is the
Mach number distribution of shocks in numerical 
simulations. 
Fig.\ref{fig:histo} shows 
the distribution of shocked cells found with the VJ method at $z=0$, 
for the total volume and in different cosmic environments.
All distributions are steep, with $\alpha \sim -3.5$ 
($\alpha = d\log N(M)/d\log M$) for the
whole volume, and $\alpha < -6$ inside the virial 
radius of galaxy clusters.
The majority of shocks are weak, with their distribution
showing everywhere a peak at $M \sim 1.5$ and a monotonic
decrease at larger $M$.

\begin{figure*}
\includegraphics[width=0.49\textwidth]{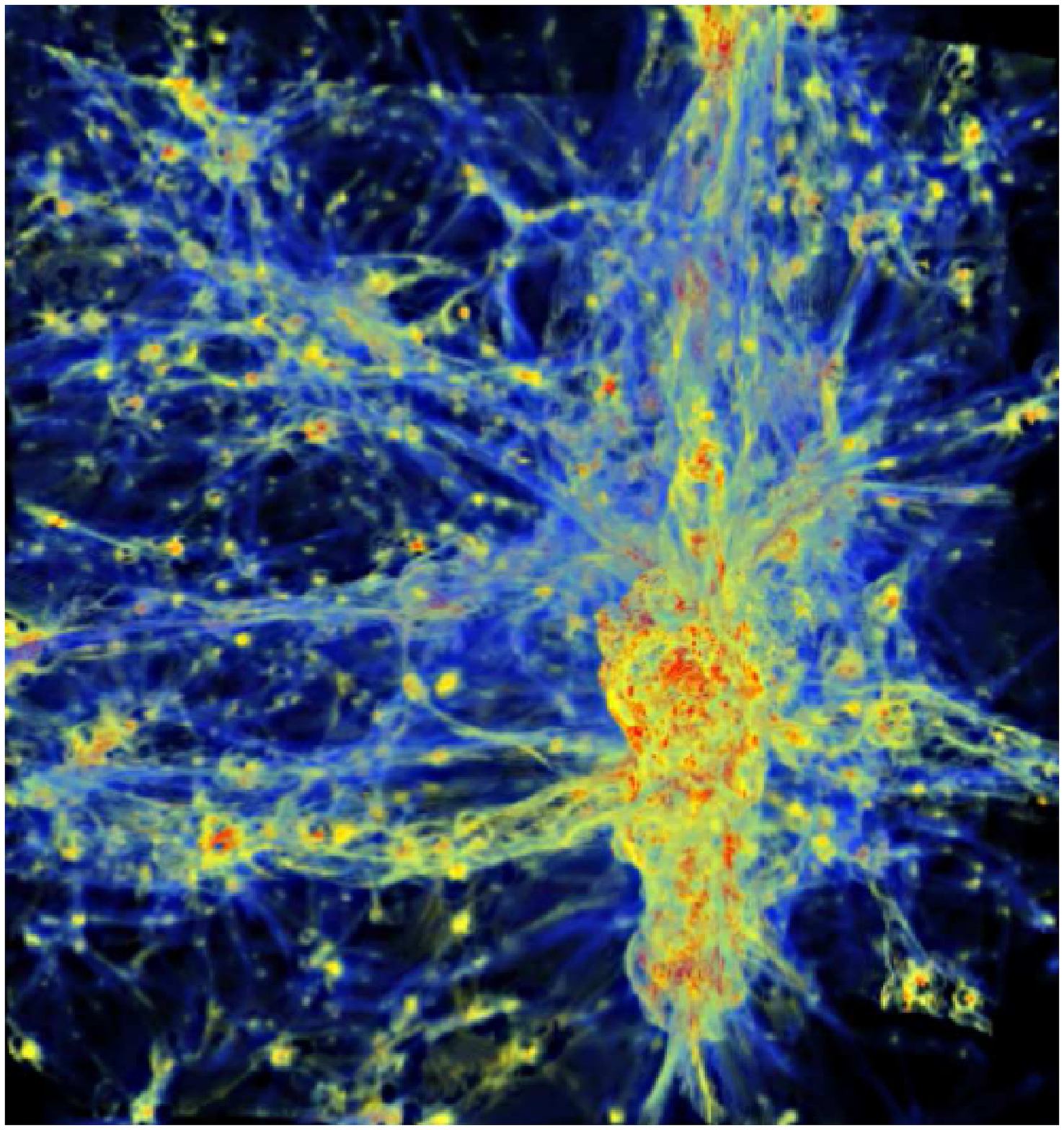}
\includegraphics[width=0.49\textwidth]{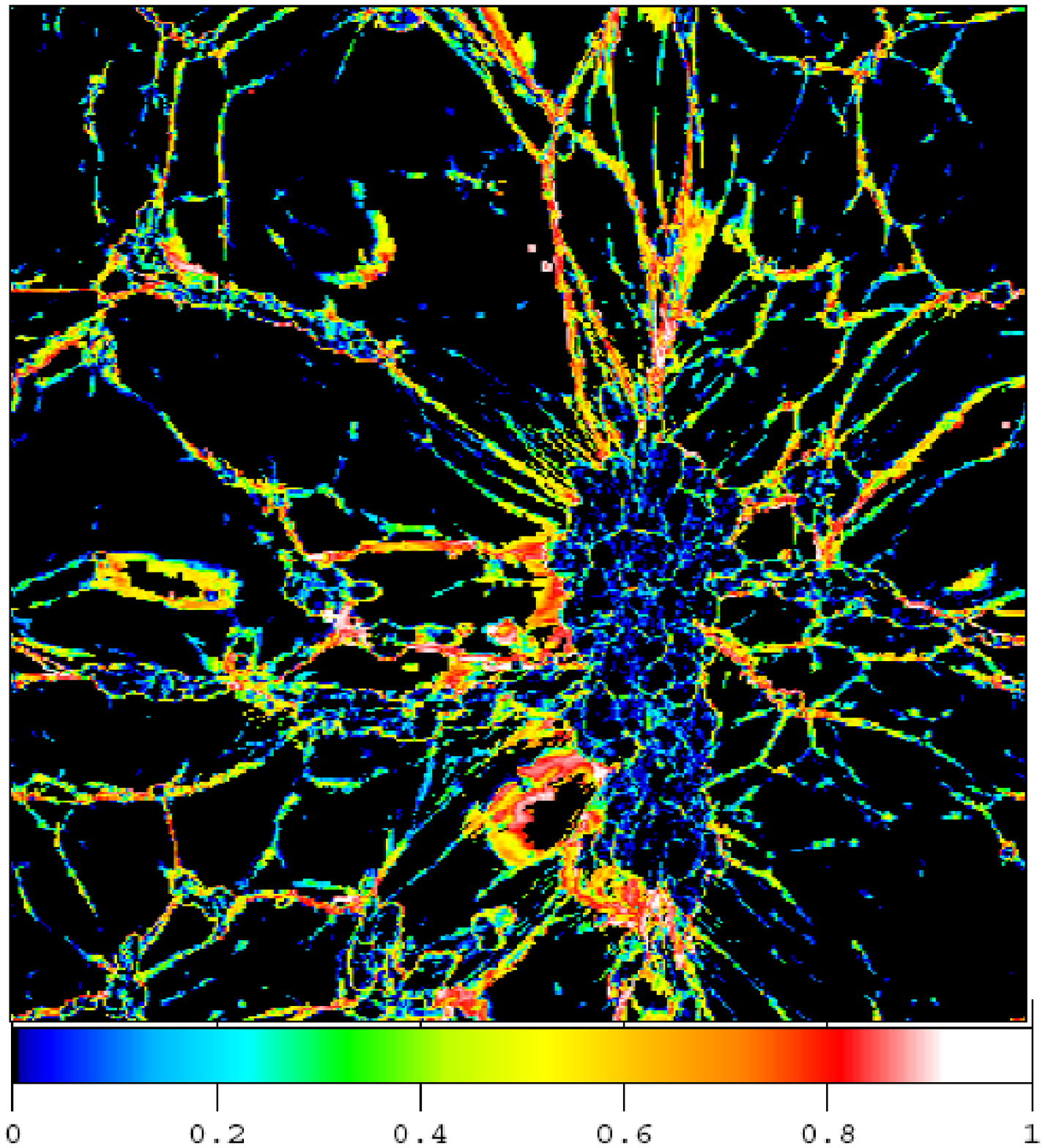}
%\includegraphics[width=0.33\textwidth]{images_new/map_cr.ps}
%;\includegraphics[width=0.33\textwidth]{images_new/map_ratio.ps}
\caption{{\it Left:}3--D rendering of the dissipated energy flux for a region of 80Mpc per side. 
Color coding goes from {\it blue} 
($f_{th}\sim 10^{33} erg/s$) to  {\it yellow} ($f_{th}\sim 10^{38} erg/s$) to
 {\it red} ($f_{th}> 10^{41} erg/s$). {\it Right:} energy ratio between 
injected CR energy flux and thermal energy flux in shock waves, for 
a slice crossing the center of the same two clusters in left panel.}
\label{fig:flux1}
\end{figure*}

\subsection{Thermal Energy Flux in Shocks.}
\label{subsec:thermal}

\begin{figure}
\includegraphics[width=0.49\textwidth]{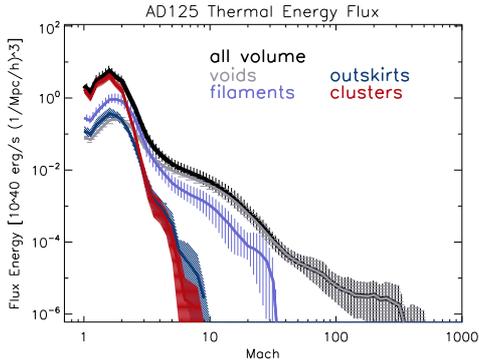}
\caption{Distribution of the thermalised energy flux 
at different overdensity bins, 
for the whole {\it AD125} and normalized to a comoving volume of $(1Mpc/h)^{3}$.}
\label{fig:histo_flux}
\end{figure}

A shock wave thermalises the post--shock region
according to Rankine--Hugoniot jumps conditions, which
relate the flux of the kinetic energy crossing 
the shock and the resulting
thermal flux in the post--shock region, $f_{th}$. 
We follow the formalism of \citet{ry03}
and \citet{ka07} to calculate $f_{th}$ at a given $M$.
 
Left panel in Fig.\ref{fig:flux1} shows a  3--D rendering of the 
thermal energy flux through shocks \footnote {The 3--D rendering 
is generated with of the visualization tool VISIVO 
(\citealt{co07}, http://visivo.cineca.it)}, while
Fig.\ref{fig:histo_flux} depicts
the distribution of $f_{th}$ at shocked cells, for 
different cosmic environments.

About $70$ per cent of the total thermal 
energy flux from shocks comes
from the virial region of galaxy clusters
and the bulk of the thermalisation is done by shocks
with $M \approx 1.8-2$. 
These relatively weak shocks are also responsible for the
thermalisation in lower density environments, but 
in these regions a sizeable fraction of the thermal energy
flux is injected at larger Mach numbers.

We find that although the total thermalised energy  
per cubic $Mpc/h$ in our simulations is consistent with previous finding,
our distribution of $f_{th}$ is steeper: we
find $\alpha_{th}\approx -3$ (with $\alpha_{th}$ taken
as $f_{th}(M)M \propto M^{\alpha_{th}}$), which should be
compared to $\alpha_{th}\approx -1.5$ to $-2$ in \citet{ry03}
and to $\alpha_{th}\approx -2.5$ in \citet{pf07}.

\subsection{Cosmic Rays acceleration.}
\label{subsec:CR}

The injection and acceleration of Cosmic Rays at shocks is a complex 
process, where several still unknown quantities play a major role (e.g. \citealt{bl04} for 
a review).
In the case of a low level of energy in form of CR, it is customary to 
describe the acceleration according to the
diffusive shock acceleration (DSA) theory (e.g. \citealt{dv81}; 
\citealt{bo78}). This theory applies when particles 
can be described by a diffusion--convection equation across the shock,
but it fails in case of strong shocks, where the pressure
in the accelerated CR back reacts on the shock itself 
(e.g., \citealt{ebj95}).

In order to have a straightforward comparison with other numerical papers on 
the issue, we estimate the ratio between the energy flux 
trough a shock and
the energy flux which is channelled into CR acceleration, $f_{CR}$,
by means of a simple parameter,
$\eta (M)=f_{CR}/f_{\phi}$ (\citealt{ka07} for an analytical expression
of $\eta (M)$), which depends
on the Mach number only. 
Fig.\ref{fig:flux1} ({it right} panel) shows 
the spatial distribution of $f_{CR}/f_{th}$ at shocks for 
a cut taken across the same region of the left panel.
The highest values of $f_{CR}/f_{th}$
are found at the interface layers of filaments or in
the outermost regions of galaxy clusters, where a substantial
population of relatively strong shocks is present. 
On the other hand the lower values
are typically found inside galaxy clusters, where the Mach number
distribution is steep and strong shocks are extremely rare; here 
$f_{CR}/f_{th} \leq 0.10$ is measured. 

As in the case of the thermal energy, we find that 
the distribution of the energy in  CR at shocks is steeper 
than that reported in other works, with the bulk of the CR injection
taking place at $M \sim 2-2.5$, at all cosmic
environments. 
Since we use an approach equivalent to that
in \citet{ry03} to evaluate the CR acceleration,
this difference is likely related to
our different shock detecting scheme, and 
expecially to our modeling
of the re-ionization process in the simulations (see VBG for
a detailed discussion).

A comparison with the results in \citet{pf06} is more
difficult since these authors use a Lagrangian Smoothed Particles Hydrodynamics
code which also include a different approach
in the calculation of CR dynamics. 
The overall distribution of the energy flux injected in CR reported
in \citet{pf06} has a slope $\alpha_{CR} \approx -1.8$ which
has to be compared with the value of $\alpha_{CR} \approx -2.2$
that we find in our simulations.

\subsection{Shocks in Galaxy Clusters.}
\label{subsec:clusters}

In this Section we studied shocks in four representative
galaxy clusters with masses 
above $5\cdot 10^{14} M_{\odot}$
and with different dynamical states: a cluster in a relaxed state,
a system with an ongoin merger with a smaller subclump, a system
approaching a major merger, and 
a post--merger system (2 Gyr after the close encounter).

\begin{figure}
\includegraphics[width=0.49\textwidth]{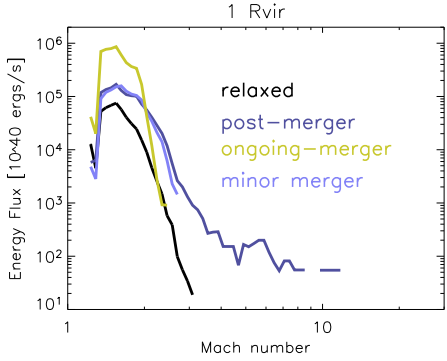}
\caption{Distribution of thermalised fluxes for the four different galaxy clusters presented in the text.} 
%Distribution are normalized for the volume
%of the most massive one, and are taken from a sphere of $2Rvir$ ({\it left}) and $1Rvir$ ({\it right}) around each galaxy cluster.} 
\label{fig:clust_distrib}
\end{figure}

Fig.\ref{fig:clust_distrib} reports
the distribution of $f_{th}$ with $M$ in shocked cells 
within $1 R_{vir}$ from
the cluster centers; the distributions in the
four clusters were normalized to the volume of the most
massive system (a sphere of radius $\sim 3Mpc$).
A very steep distribution of the energy flux trough shocks
is found in the relaxed and in the minor merger case, while 
in the case of the ongoing merger and in the post merger
case the distributions also show tail of higher Mach numbers.
Inside $R_{vir}$ no shocks with $M > 3$
are detected, except for a few in the case of the post--merger
system, and this is in line with X--ray
observations of real merging clusters (e.g. \citealt{mv07}). 
We remark that our findings are in line with
expectations from semi--analytical treatment of shocks in virialzied merging galaxy
clusters \citep{gb03}.
On the other hand, we find significantly less strong shocks than in 
\citet{pf07}. In this case the differences are due to the different 
numerical scheme and to the procedure adopted to characterise shocks, and 
highlight the level of present ucertainty in this issue. 

\begin{figure}
\includegraphics[width=0.49\textwidth]{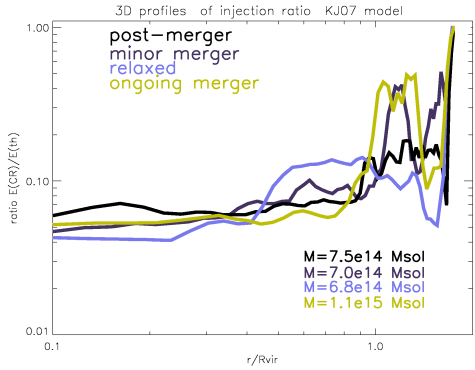}
\caption{Volume averaged profiles of the injection
efficency, $f_{CR}/f_{th}$, for the four galaxy clusters.} 
\label{fig:prof_cr_clusters}
\end{figure}

The radial behaviour of the $f_{CR}/f_{th}$ for these four clusters
is reported in Fig.\ref{fig:prof_cr_clusters}, where we adopted 
the \citet{ka07} formulation.
Inside the virial radius we do not find any relevant difference 
among our clusters, and an average value of $f_{CR}/f_{th}<10$ is found.
This is because shocks crossing the innermost
regions are weak, independently of the dynamical status
of the clusters. 
Although we do not follow the advection of CR with cosmic time, and this
makes a comparison with observations more challenging, our estimates
appear in line with existing upper limits from Radio observations
for $\mu G$ magnetic field in galaxy clusters \citep{gb07}, as in our
case the spectrum of CR is steeper than previous work (with an average 
slope of $\delta \sim -3.5$, see VBG for a detailed discussion).

\section{Discussion and Conclusion.}
\label{sec:conclusions}

In this paper we have investigated the
properties of shock waves
in LSS simulations produced with the cosmological code ENZO 
and by means of a new
detection scheme to measure the shocks, based on
velocity jumps.

In the following we summarize the main results obtained:

\begin{itemize}
 
\item we detect morphologies of LSS shock--patterns which are qualitatively
in agreement with previous numerical works (e.g. \citealt{mi01}), with
strong shocks enveloping filaments and sheets of matter, and weaker shocks
hosted inside galaxy clusters.

\item We measure the number distribution of shocks as a function of the
Mach in all cosmological environments. The bulk of cosmological shocks 
is made by weak $M \leq 2$ shocks and their distribution can be grossly described by a steep power law $N(M) \propto M^{\alpha}$, with
$\alpha \approx -3.5 $. In the case of galaxy clusters 
$\alpha \approx -6$, demonstrating the increasing rarity 
of strong shocks in these denser
(and hotter) regions.

\item Following \citet{ry03} and \citet{ka07} we 
calculate the energy rate dissipated
in thermal energy at shocks, finding that roughly a 
70 percent of the thermal dissipation
in the whole volume happens inside galaxy clusters, at an average
Mach number of $M \approx 2$.
Although in qualitative agreement with previous studies,
the energy distributions we measure in all environments are
steeper than those obtained by \citet{ry03} and by \citet{pf06}.

\item We calculate the efficeny of CR acceleration
for our simulations. Also in this case our results are in
qualitatively in line with previous findings,
although our energy distributions with $M$ are sistematically 
steeper than those in \citet{ry03} and
slightly steeper than those in \citet{pf06}.

\item We report on the properties of shocked cells 
propagating in four representative galaxy
clusters of our sample. The average Mach number of shocks 
within $1 R_{vir}$ is $M \approx 1.5$
and this is in line with semi--analytical studies dealing with
mergers of virialised systems \citep{gb03}. 
Also, the rariry of stronger shocks ($M > 2-3$) is found in
line with the the rarity of shocks detected so far by X--ray
observations (e.g. \citealt{mv07}).

\end{itemize}

\bigskip
F. V. thanks K. Dolag, D. Ryu, H. Kang, G. Tormen, L. Moscardini, S. Giacintucci and R. Brunino 
for useful discussions and helps. 
We acknowledge partial support through grant ASI-INAF I/088/06/0, and the
usage of computational resources under the CINECA-INAF agreement.

%% ======================================================================= %%
%% REFLIST REFLIST REFLIST REFLIST REFLIST REFLIST REFLIST REFLIST REFLIST %%
%% ======================================================================= %%

\bibliographystyle{aa.bst}
\bibliography{vazza_1.bib}

\end{document}